\documentclass[12pt,a4paper]{article}%
\usepackage{amsfonts}
\usepackage{amsmath}
\usepackage{geometry}
\usepackage{amssymb}
\usepackage{graphicx}%
\providecommand{\U}[1]{\protect\rule{.1in}{.1in}}

\begin{document}

\title{Determinism and the Theory of Every Thing}
\author{M.B. Altaie\thanks{email: maltaie@yu.edu.jo}\\Department of Physics, Yarmouk University, 21163 Irbid, Jordan.}
\maketitle

\begin{abstract}
Recently Gerard 't Hooft proposed a structure for a universe overwhelmed with
a control by a Theory of Everything (arXiv:1709.02874). He concludes, among
many other things, that such a universe could be fully deterministic and that,
accordingly, the divine intervention will be eliminated. Here I discuss such a
possibility and show that a fully deterministic universe will turn out to
become the divine himself, thus verifying the consistency of Einstein's belief.

\end{abstract}

\section{Introduction}

The present status of theoretical physics, which is something like a
stalemate\ is provoking physicists to speculate about the possibilities of
getting out of this dilemma. The incompatibility of quantum mechanics and
general relativity on the theoretical level stands as a firewall that cannot
be breached, whereas any further major development in black hole physics,
particle physics, and cosmology would require solving this enigma. Several
proposals are on the table but none seem to work. It was claimed more than
three decades ago that we are at the verge of a Theory of Everything which may
solve many problems of contemporary physics \cite{Ellis}, but the years have
shown that we are still far a way from achieving such a goal. The problem seem
to be more fundamental than we thought, and as 't Hooft beautifully remarks
"\emph{we are just baboons who have only barely arrived at the scene of
science}" \cite{Hooft1}.

The problem of understanding the universe and the role played by the laws of
nature in controlling events and in driving the phenomena taking place in this
universe goes beyond the task of science at this stage of the development of
our mind. Philosophy and religion may provide some help to widen our scope in
looking for a solution, however, not many physicist have faith in
philosophical methods and very few of them believe that religion can inspire
anything at all in science. This attitude may have resulted from the
historical experience gained during the last two centuries or so and from the
naive personal figure of God offered by religions. Nevertheless one may find
that the pioneering generation of the seventeenth century had much more
faith-inspired motivations to view the world. In this respect we find Leibniz
quoted saying "\emph{the imaginary numbers are a wonderful flight of God's
spirit; they are almost an amphibian between being and not being}"
\cite{Leibniz}. Indeed we may need to extend our vision so as to cover the
wider range of existence which include the imaginary space and the space-like
universes on the same footing as we have considered the multi-dimensional
space-time. The point that may cause some disturbance is that our attitude
could be the question of the objectivity of such entities, nevertheless it
remains a fact that such entities like imaginary numbers, though not
measurable, are part of our mathematical description of our world playing an
essential role in our world of electromagnetic radiation and quantum tunneling
(a purely real quantum state will not tunnel through potentials).

Resolving the present big problems in physics, like quantum gravity, requires
a new perspective in our mathematical formulations as well as the physical
realization of some basic concepts. We should remind ourselves that we have no
consensus as of yet on interpreting the problem of measurement in quantum
mechanics. String theory might appear to be promising on the conceptual level,
but the formal treatments of the strings through differential calculus by
adopting the assumption of spacetime continuum, hinders the breaching into new
area and, in most cases, aborts generating new predictions.

\section{Setting the Theater}

In an article entitled "Free will in the Theory of Everything" \cite{Hooft1}
Professor 't Hooft imagined a scenario of thought by which he placed a being
playing the role for running the universe. He identified some basic
\emph{demands} for such a God helping reduce his duties and making it more
efficient. However, it \ is not clear whether such a being himself is going to
formulate the laws of Nature or if such laws are to be given by his computer
scientists and mathematicians. Beside this, there is no clear distinction
between the Laws of Nature and Laws of Physics, an issue which I have
discussed in a pervious work \cite{Basil}. The basic problem in this context
is the sustainment of the universe, and the question remains will be whether
God is directly intervening in every motion of every particle in the universe
or should he do so by delegating secondary agents? 't Hooft finds that God has
chosen to adopt certain laws of Nature that makes his work in sustaining the
universe more efficient. The algorithms by which the computers work are some
basic demand and a set of rules.

\subsection{Limitations}

There are several embedded limitations in the proposed model for the universe
by 't Hooft, such as

\begin{enumerate}
\item The type of the imposed causality makes this universe time-like, thus
may deny the possibility that some space-like events might contribute to the
event happing in our time-like universe.

\item It is required that spacetime is continuous and local, a requirement
which prevent the possibility of having intrinsically quantized spacetime,
beside the fact that non-locality is a natural requirement of quantum mechanics.

\item The universe is superdeterministic, an idea that remains to be absurd no
matter how it is defended.

\item The rule of the imposed speed limit leaves no room to deal with
space-like patches of the spacetime through which we can explain certain
phenomena of our time-like universe (e.g. magnetic monopoles and the tachyonic
behavior of some events). Instead one may require the presence of the
principle of invariance, out of which many sub-rules could naturally emerge.
\end{enumerate}

Despite that "\emph{it may lead exactly to our universe}" as 't Hooft is
suggesting, such limitations will captivate our mind to work within the rules
of the present game of knowledge about the universe. This is certainly
something which is not expected to lead us to the Theory of Everything.

\subsection{Alternatives}

Some alternative rules to those suggested by 't Hooft might be proposed. These
might help open the way for a new realization of the world around us and help
in formulating better mathematical as well as physical perspective for the world.

\begin{enumerate}
\item Causality is an apparent relationship between the variables involved in
an event or a series of events. Causality is not a single attribute but a
process indicating chronological order as well as a result.

\item Locality may not be a good choice to be a rule, instead we may require
the principle of invariance as a good demand by which laws, specifically
conservation laws, become the product of the invariance under certain
symmetries. This applies to Rule \#2 of 't Hooft concerning the ultimate
velocity. It is actually a presentation of the non-locality and the invariance
of spacetime under translation.

\item It might be of help to assume that all physical entities in the world,
including spacetime, are atomistic and are under continued re-creation process
\cite{Altaie1}. This may resolve the basic problem of quantum measurement and
explain many phenomena like quantum entanglement and the quantum Zeno effect.

\item The universe is indeterministic, as viewed by us, but obeys certain set
of rules (laws) which help making it comprehensible and predictable.
\end{enumerate}

\section{Indeterminism and the Hidden Variable}

The indeterminism of quantum phenomena is a well established experimental
fact, however the orthodox interpretation of such a fact is not yet well
accepted as many physicists including 't Hooft himself, are not happy with the
Copenhagen interpretation \cite{'t Hooft2}. Among the other alternatives to
the Copenhagen wavefunction collapse interpretation comes the suggestion by
David Bohm of some sort of hidden variable that may come at the play. Such a
hidden variable could present the case of quantum measurement to become
deterministic according to Bohm's scheme \cite{Bohm}. Again very few
physicists believe that such an approach would be successful in adding
anything to quantum physics beside that many of them where stunned by John
Bell's theorem ruling out the local hidden variable theories \cite{Bell}.
However, in this respect there remains the question whether quantum mechanics
is an ensemble theory that is reflecting a statistical nature of the
microscopic world or is it suitable for applications to a single particle
case. This is the essence of the measurement problem in quantum mechanics.

One approach for understanding quantum states is to notice that it is
generally an ever-changing state, a feature which was originally known for the
wave-mechanical description, and in fact this is one reason for the success of
wave-mechanical description of the quantum states. As such is the state of a
particle, one may alternatively suggest that the quantum states are under
continued re-creation allowing them to be regenerated anew in each shot. One
may further assume that the re-creation rate is proportional to the energy
content of the system, perhaps it might be plausible to suggest that the
frequency of re-creation may be calculated from the basic form
\begin{equation}
f=\frac{E}{h},
\end{equation}
where $h$ is Planck's constant. It is clear then that systems with high energy
content will have high rate of re-creation. An electron at rest will have a
rate of re-creation of a bout $1.\,\allowbreak23\times10^{20}$\ s$^{-1}$. With
this assumption of re-creation we can glimpse that there will be some inherent
uncertainty in measuring any pair of complementary observables of the system.
For example if the position is re-created then an infinitesimal shift in the
position is expected, which correspond to the generation of momentum. Then,
upon re-creating the momentum an inherent generation of position would take
place. If we demand that the system to develop unitarily, that is to say that
the state of the system remains the same, up to a phase factor, then
\begin{equation}
\left(  x\frac{\partial}{\partial x}-\frac{\partial}{\partial x}x\right)  =1,
\end{equation}
using the definition of the momentum operator, $p=-i\hslash\frac{\partial
}{\partial x},$ this would imply that
\begin{equation}
\left[  x,p\right]  =i\hslash
\end{equation}
Accordingly, the uncertainty relationship between position and momentum is
established. This proposal of re-creation has many implication shown in a
previous work \cite{Altaie1}.

\section{Determinism and the Divine Intervention}

Does determinism lead to the elimination of the divine intervention? Most
people think that divine intervention is needed only upon the creation of the
universe, an event after which the creator became redundant. They think that
laws of Nature can stand alone and can perform the world's phenomena taking
place without the need for any other agency. This, in fact, is a sort of
unsubstantiated assumption normally embodied in our argumentation. In fact we
need to prove that laws of Nature can act on their own before using such an
assumption. If we agree to define the laws of Nature as being those events
happening regularly without specifying any cause or explanation for it, then
we would recognize the fact that the laws of nature are the regular phenomena
taking place as it happens. But quantum mechanic tells us that such phenomena
are generally indeterministc. This means that there should be some kind of an
agency playing with the probabilities and so controlling the phenomena. Such
an agency cannot be thought of as being part of the world, because in such a
case it has to abide by the basic rules of indeterminism again and so will
need a another agency to drive it. This would be compelling reason to assume
that if such an agency would ever exist then it has to be something beyond our
physical world.

Once we try to explain any natural phenomena we are devising a Law of Physics.
These laws are our own imagination, our own comprehension of the phenomena.
The puzzle is that our understanding of the laws of Nature appears to follow
certain logical structure we call mathematics, which was shown over the ages
to have a strong deductive power. Since then we thought that the laws of
physics are not only descriptive of Nature, but also are prescriptive.

By the fact that laws of nature are probabilistic we can confidently suggest
that no self-driven system can be achieved. So, if one thinks that a
deterministic universe would leave God redundant, he need to cover the
incompleteness suggested by G\H{o}del's theorem, beside the need for a driver
to fuel the dead equations. Taking into consideration the suggestions of 't
Hooft we conclude that the deterministic universe will turn to be the divine
himself, a conclusion which turns to be in agreement with Einstein's concept
of God; standing for all the order in the universe.

\end{document}